\documentclass[12pt,letterpaper]{article}

\usepackage{amsmath,amssymb,calc}
\usepackage{graphicx}
\usepackage{color}

\newcommand\be{\begin{equation}}
\newcommand\ee{\end{equation}}
\newcommand{\bea}{\begin{eqnarray}}
\newcommand{\eea}{\end{eqnarray}}

\newcommand{\nn}{\nonumber}
\newcommand{\pd}{\partial}

\def\id{\protect{{1 \kern-.28em {\rm l}}}}

\def\1{^{(1)}}
\def\0{^{(0)}}
\def\2{^{(2)}}

\def\id{\protect{{1 \kern-.28em {\rm l}}}}

\let\non\nonumber

\begin{document}
\begin{titlepage}
\begin{center}
\hfill \\

\vskip 1cm
{\Large \bf Scalar Mesons in Holographic\\ \vspace{0.4cm} Walking Technicolor}

\vskip 1.5 cm
{\bf  Lilia Anguelova${}^a$\footnote{languelova@perimeterinstitute.ca}, Peter Suranyi${}^b$\footnote{peter.suranyi@gmail.com} and L.C.R. Wijewardhana${}^b$\footnote{rohana.wijewardhana@gmail.com}\\
\vskip 0.5cm  {\it ${}^a$ Perimeter Institute for Theoretical Physics, Waterloo, ON N2L 2Y5, Canada\\ ${}^b$ Dept. $\!$of Physics, University of Cincinnati,
Cincinnati, OH 45221, USA}\non\\}

\end{center}
\vskip 2 cm
\begin{abstract}
We study the spectrum of scalar mesons in the holographic dual of walking technicolor, obtained by embedding D7-$\overline{{\rm D}7}$ probe branes in a certain type IIB background. The scalar mesons arise from fluctuations of the probe techniflavour branes and complement the (axial-)vector meson spectra that we investigated in earlier work. By explicitly finding the spectrum of scalar masses, we show that the nonsupersymmetric D7-$\overline{{\rm D}7}$ embedding is stable with respect to such fluctuations. Interestingly, it turns out that the mass splitting between the scalar and vector meson spectra is of subleading order in a small parameter expansion. It is noteworthy that this near-degeneracy
may not be
entirely due to a small amount of supersymmetry breaking and thus
could indicate the presence of some other (approximate) symmetry in the problem.
\end{abstract}
\end{titlepage}

\tableofcontents

\section{Introduction}

Gauge symmetries play a crucial role in modern particle physics. However, they also imply massless gauge bosons. To reconcile gauge invariance with massive gauge  interaction mediators, one invokes spontaneous symmetry breaking. This is precisely the case with the electroweak interactions of the Standard Model of particle physics. Unfortunately though, the origin of the symmetry breaking has not yet been established experimentally. A leading candidate for the underlying mechanism is
the condensation of a weakly coupled fundamental scalar field,
whose manifestation in nature would be the existence of the Higgs boson. There has been a lot of excitement recently about a possible discovery of a spin-zero particle at the LHC, which might be the Higgs. However, an actual confirmation requires more data. Furthermore, there are other possible explanations for the existence of a (fundamental or composite) scalar with mass in the relevant energy range. Regardless of the experimental status of the Higgs boson, however, the existence of a fundamental scalar would lead to well-know theoretical problems. This motivates the search for alternative explanations of the origin of mass in the Standard Model.

An appealing alternative relies on dynamical effects as the cause of symmetry breaking. In this approach, the role of the Higgs is played by a composite scalar, formed as a result of strongly coupled gauge dynamics. Theories of that kind are known as technicolor theories \cite{Tech}. In fact, to be phenomenologically viable, they have to be characterized by a gauge coupling that varies very slowly in an intermediate energy range. This phenomenon is called walking, in analogy with the more familiar running of couplings. Walking technicolor was first proposed in \cite{WalkTech}. There has been a lot of work on this subject over the years. However, a major hindrance for progress has been the need to study strongly coupled gauge theories. This is clearly, beyond standard perturbative QFT methods. As a result, the computation of electroweak observables from models of dynamical electroweak symmetry breaking has long been a challenge.

In recent years, a new theoretical tool was developed, that addresses precisely the investigation of the non-perturbative regime of gauge theories. This is the gauge/gravity duality, which allows one to study the strongly coupled regime of a gauge theory via a weakly coupled (called dual) gravitational background in a different number of dimensions. Thus non-perturbative QFT problems are mapped to (almost) classical gravitational computations. This powerful method has already produced interesting insights into a variety of phenomena, ranging from hydrodynamics \cite{hydro} to superconductivity \cite{scond}.

Of, perhaps, greatest interest for particle physics is the gravity dual, studied by Sakai and Sugimoto \cite{SS}. This is a dual of a gauge theory that captures the characteristic features of QCD, including chiral symmetry breaking. It arises from a certain D8-$\overline{{\rm D}8}$ brane configuration in a background sourced by a stack of D4 branes in type IIA string theory. The Sakai-Sugimoto model, as well as a wide variety of other similar D-brane configurations in both type IIA and type IIB, have been easily adapted to study QCD-like technicolor \cite{HolTech}.\footnote{We should also note the existence of a huge literature, loosely inspired by AdS/CFT, on bottom-up holographic technicolor models \cite{PhenoHolTech}, which do not arise from any D-brane construction.} In these models one can compute explicitly the $S$ parameter, which is an important electroweak observable \cite{PT}. However, as is to be expected, the answer is not compatible with the known experimental bounds.

In \cite{LA,ASW}, we studied the gravity dual of a model of walking technicolor. This model has two key ingredients. The first is a certain type IIB background \cite{NPP} sourced by D5 branes wrapped on an $S^2$, whose dual is a walking gauge theory. The second ingredient, which introduces the techniflavor degrees of freedom, is a particular D7-$\overline{{\rm D}7}$ probe-branes embedding \cite{LA} in the background of \cite{NPP}. In \cite{ASW}, we showed that the $S$ parameter of this model can be small enough to be compatible with experiment.\footnote{It is, perhaps, worth pointing out that the recent work \cite{LRTZ} finds a lower bound on the $S$ parameter in a large class of bottom-up holographic technicolor models, which is in conflict with observation. However, the considerations in the bottom-up approach rely on certain assumptions, that are not satisfied in the actual supergravity duals arising from appropriate D-brane configurations. Thus, the results of \cite{LRTZ} do not apply to our model.} This demonstrates, by means of an example, that walking technicolor models can, indeed, satisfy the $S$ parameter test for phenomenological viability.

Here we continue the study of the model of \cite{LA,ASW}. In addition to the spectra of vector and axial-vector mesons, that were investigated in \cite{ASW}, there should also be scalar mesons. The latter arise from the fluctuations of the techniflavor D7-$\overline{{\rm D}7}$ probe branes.
As there are two directions transverse to the probe worldvolume, there are two kinds of scalar mesons.
Since the D7-$\overline{{\rm D}7}$ embedding is not supersymmetric, it is an open question whether the scalar spectrum will contain any tachyonic modes or not. In this paper we show that the embedding is in fact stable with respect to such fluctuations. Namely, we find the mass spectrum of scalar mesons explicitly and thus demonstrate that all mass-squareds are positive. Interestingly, it turns out that the scalar meson spectrum is only slightly shifted compared to the spectrum of vector mesons.\footnote{By that we mean that it is shifted only by terms that are subleading in an expansion in a certain small parameter.} This is not a priori clear,
as examplified by the models of \cite{KS,SS} in which the scalar and vector meson spectra differ by order one terms. The near-degeneracy in our case could indicate that the amount of supersymmetry breaking, induced by our D7-$\overline{{\rm D}7}$ embedding, is small. Recall that a massive ${\cal N} = 1$ vector multiplet has the same field content as a massless vector multiplet and a chiral multiplet. Thus, the massive vector and scalar states belong to the same long ${\cal N}=1$ representation. However, in the present case, the two kinds of scalar mesons turn out to have slightly different mass shifts compared to the vector meson spectrum. This suggests that there may be some
additional mechanism at work,
which would be very interesting to understand in the future.

We should point out that the scalar spectrum investigated here is not directly relevant for settling the interesting open question of whether there is a technidilaton. The latter is a light composite scalar\footnote{By light we mean with mass that is (parametrically) much smaller than the masses of the other states in the spectrum (in particular, the technirho meson).}, which is
the would be pseudo-Goldstone boson associated with the spontaneous breaking of
near-conformal invariance of the walking regime. In models, in which the breaking of conformal invariance is entirely due to the flavor probes (as in \cite{KS,KLP}, for example), the technidilaton (if present) should appear in the spectrum of scalar mesons arising from the fluctuations of those probes. Such mesons are exactly the kind of scalar states studied here. However, in our model the technicolor background already encodes conformal symmetry breaking. Hence, in our case the technidilaton should arise predominantly from the technicolor, not the techniflavor, sector.\footnote{It is rather likely that this state will turn out to be related to the light scalar mode found in \cite{ENP}.} Thus the question of whether our model has a technidilaton is beyond the scope of the present study. We will address this problem in another publication.

The present paper is organized as follows. In Section 2, we briefly recall the basic ingredients of the walking technicolor model of \cite{LA,ASW}. We also derive the action for the two kinds of fluctuations of the flavour probe branes, that give rise to the scalar meson spectrum. In Sections 3 and 4 we study the mass spectra of each of the two kinds of fluctuations in detail. More precisely, we show that to leading order in a small parameter expansion they are the same as the vector meson spectrum of \cite{ASW}. Furthermore, we compute the first subleading correction, that gives the mass shift between the scalar and vector boson spectra. Finally, Appendices A and B contain some technical details of the computations in the main text.

\section{Action for scalar mesons}
\setcounter{equation}{0}

In this section we review the basic ingredients of the technicolor model that we will be studying. More precisely, we recall the background that encodes the technicolor degrees of freedom, as well as the D7-$\overline{{\rm D}7}$ probe-branes embedding that introduces the techniflavour ones. Furthermore, we derive the action for the fluctuations of the D7-$\overline{{\rm D}7}$ embedding, which give rise to the spectrum of scalar mesons in this model.

\subsection{Background and flavour probes}

The background we will consider is due to a stack of D5 branes wrapped on an $S^2$.
It consists of nontrivial ten-dimensional metric, RR 3-form flux and string dilaton.
Furthermore, it is a solution of the type IIB equations of motion with ${\cal N}=1$ supersymmetry.
In \cite{NPP}, this solution was found as an expansion in a small parameter. For our purposes, it is enough to consider only the leading order contributions to the background fields. Note that, to this order, the string dilaton is constant. The leading order metric is \cite{NPP}:
\bea \label{BM}
ds^2 &=& A \left[ dx_{1,3}^2 + \frac{cP_1' (\rho)}{8} \left( 4 d\rho^2 + (\omega_3 + \tilde{\omega}_3)^2 \right) \right. \nn \\
&+& \left. \frac{c\,P_1 (\rho) \coth (2 \rho)}{4} \left( d\Omega_2^2 + d\tilde{\Omega}_2^2 + \frac{2}{\cosh ( 2\rho)} (\omega_1 \tilde{\omega}_1 - \omega_2 \tilde{\omega}_2) \right) \right] \, ,
\eea
where
\be \label{AP1}
A=\left( \frac{3}{c^3 \sin^3 \alpha} \right)^{1/4} , \quad  P_1' (\rho) =\frac{\pd P_1 (\rho)}{\pd \rho} \,\, , \quad P_1 (\rho) = \left( \cos^3 \alpha + \sin^3 \alpha \left( \sinh (4 \rho) - 4 \rho \right) \right)^{1/3} \, ,
\ee
\bea
\tilde{\omega}_1 &=& \cos \psi d\tilde{\theta} + \sin \psi \sin \tilde{\theta} d \tilde{\varphi} \,\, , \hspace*{2cm} \omega_1 = d \theta \,\, , \nn \\
\tilde{\omega}_2 &=& - \sin \psi d\tilde{\theta} + \cos \psi \sin \tilde{\theta} d \tilde{\varphi} \,\, , \hspace*{1.6cm} \omega_2 = \sin \theta  d \varphi \,\, , \nn \\
\tilde{\omega}_3 &=& d \psi + \cos \tilde{\theta} d \tilde{\varphi} \,\, , \hspace*{3.7cm} \omega_3 = \cos \theta d \varphi
\eea
and
\be
d\tilde{\Omega}_2^2 = \tilde{\omega}_1^2 + \tilde{\omega}_2^2 \,\, , \hspace*{2cm} d\Omega_2^2 = \omega_1^2 + \omega_2^2 = d \theta^2 + \sin^2 \theta d \varphi^2 \,\, .
\ee
The constants $c$ and $\alpha$ above are parameters of the solution and, as in \cite{NPP}, we have set $\alpha' = 1 = g_s$. For an explicit expression for the RR flux, see \cite{NPP}.\footnote{Note also that the background described above is the leading order solution in an expansion in $1/c$ with $c>\!\!>1$.}

This background is dual to a color gauge theory with a coupling constant, which exhibits walking behavior. Namely, in a certain intermediate energy range, the gauge coupling is almost constant as a function of the energy. In this walking region, $\rho$ is always of order 1 or larger \cite{NPP}. Hence, $\coth(2\rho) \approx 1$ while $\frac{1}{\cosh(2\rho)}$ is negligible. In addition, the walking region is characterized by \cite{NPP}:
\be
\beta \equiv \sin^3 \alpha <\!\!< 1 \, .
\ee
As a result, to leading order in small $\beta$, one can use the approximations:
\be
P_1 = 1 \qquad , \qquad P_1' = \frac{2}{3} \beta e^{4 \rho} \, .
\ee
Therefore, the metric (\ref{BM}) simplifies to:
\be \label{leadMetric}
ds^2_{{\rm walk}} = A \left[ \eta_{\mu \nu} dx^{\mu} dx^{\nu} + \frac{c}{12} \beta e^{4 \rho} \left( 4 d\rho^2 + \left( \omega_3 + \tilde{\omega}_3 \right)^2 \right) + \frac{c}{4} \left( d \Omega_2^2 + d \tilde{\Omega}_2^2 \right) \right] .
\ee

To add flavor degrees of freedom, one can introduce D7 probe branes in the above background. In fact, as shown in \cite{LA}, there is an embedding of D7 and anti-D7 branes such that the D7 and $\overline{{\rm D}7}$ merge at some finite value, $\rho_0$, of the radial coordinate. This realizes geometrically chiral symmetry breaking and thus enables us to construct a model of walking technicolor. In \cite{ASW}, we studied the spectra of vector and axial-vector mesons in this model. Here we will investigate the spectrum of scalar mesons. The latter arise from the fluctuations of the ${\rm D}7$-$\overline{{\rm D}7}$ embedding. So let us, first, describe this embedding in more detail.

The eight worldvolume directions of the probe branes are $x^{\mu}$, $\rho$, $\psi$, $\tilde{\theta}$ and $\tilde{\varphi}$. The two transverse space coordinates are functions of $\rho$ only: $\theta = \theta (\rho)$, $\varphi = \varphi (\rho)$. The equations of motion for $\theta (\rho)$ and $\varphi (\rho)$, that arise from the DBI action for the D7 branes, can be solved by the following classical configuration \cite{LA,ASW}:
\be \label{clSol}
\theta_{cl} = \frac{\pi}{2} \qquad {\rm and} \qquad \varphi_{\rho}^{cl} = \pm 2 \sqrt{\frac{\beta}{3}} \,e^{2\rho_0} \frac{1}{\left(1 - e^{4(\rho_0 - \rho)} \right)^{1/2}} \,\,\, ,
\ee
where $\varphi_{\rho}^{cl} \equiv \frac{\pd \varphi^{cl}}{\pd \rho}$.\footnote{Of course, we can integrate $\varphi_{\rho} (\rho)$ to obtain $\varphi (\rho)$. However, this will not be necessary or useful, since the function $\varphi (\rho)$ will appear in the relevant action below only via its derivative.} For future use, let us also introduce a new, worldvolume, radial coordinate:
\be \label{ChVar}
z = \pm \sqrt{e^{4 (\rho - \rho_0) } - 1} \, ,
\ee
which runs over both the ${\rm D}7$ and $\overline{{\rm D}7}$ branes and is such that $z=0$ when $\rho = \rho_0$. Note that, up to an overall constant, this is the same worldvolume variable as the one in equation (4.2) of \cite{ASW}. In terms of $z$, the classical solution for the derivative of $\varphi$ is:
\be \label{derVarPhi}
\varphi_z^{cl} = \frac{\pd \rho}{\pd z} \,\,\varphi_{\rho}^{cl} = e^{2\rho_0} \sqrt{\frac{\beta}{3}} \,\frac{1}{\sqrt{1+z^2}} \,\, .
\ee
Note also that at $z=0$ the above expression for $\varphi_z^{cl}$ is regular, unlike $\varphi_{\rho}^{cl}$ in (\ref{clSol}) at $\rho = \rho_0$. This is a key reason to introduce the variable $z$. Namely, since we would like to expand the field $\varphi$ in fluctuations around the configuration $\varphi_{cl}$, an appropriate choice of coordinate should be such that $\varphi_{cl}$ does not diverge anywhere along its profile.

\subsection{Fluctuations of ${\rm D}7$-$\overline{{\rm D}7}$ embedding}

To find the spectrum of scalar mesons in our model, we want to study 4d spacetime-dependent fluctuations of the ${\rm D}7$-$\overline{{\rm D}7}$ embedding around the configuration (\ref{clSol}). So we promote the embedding functions to fields in spacetime and expand them as:
\be \label{tpFluct}
\theta (z,x^{\mu}) = \theta_{cl} + \delta \theta (z,x^{\mu}) \quad , \quad \varphi (z,x^{\mu}) = \varphi_{cl} (z) + \delta \varphi (z,x^{\mu}) \,\, .
\ee
To obtain the action for the fluctuations $\delta \theta$ and $\delta \varphi$, we need to compute
\be
{\cal L}_{DBI}^{{\rm D}7} = - \sqrt{-\det g_{8d}} \,\, ,
\ee
where $g_{8d}$ is the induced metric on the D7 worldvolume, and expand it to second order in fluctuations.\footnote{We have dropped an overall constant, that is due to the constant string dilaton in this background.} To do that, we substitute
\bea\label{d1}
d\theta = \theta_z dz + \theta_{\mu} dx^\mu \quad , \quad d\varphi = \varphi_z dz + \varphi_{\mu} dx^\mu
\eea
in (\ref{leadMetric}) with the change of variable (\ref{ChVar}) performed. Note that $\theta_z$, $\theta_{\mu}$ and $\varphi_{\mu}$ are all first order in the fluctuations, since $\theta_{cl} = const$ and $\pd_{\mu} \varphi_{cl} = 0$. However, $\varphi_z$ does contain a zeroth order contribution from $\varphi_{cl}$. The details of the computation of $\sqrt{-\det g_{8d}}$ can be found in Appendix A. The resulting Lagrangian is the following:
\be\label{Lz}
{\cal L} = const \sqrt{3(1+z^2) \!\left[4\,\theta_z^2+\sin^2 \!\theta \left(4+c\,\theta_\mu^2\right)\varphi_z^2\right]+e^{4 \rho_0}\beta\,z^2\left(4+c\,\theta_\mu^2+c\,\varphi_\mu^2\right)} \, .
\ee
To study the action for the fluctuations to quadratic order, we need to expand the last expression further. First, however, note that all terms under the square root are quadratic in the fields and there is no mixing between the $\theta$ and $\varphi$ contributions to second order; in particular $\sin^2 \!\theta = \sin^2 \!\left( \frac{\pi}{2} + \delta \theta \right) \approx 1 - (\delta \theta)^2$. Hence, the desired action for each of the two fields can be obtained by first turning off the fluctuations of the other field and then expanding to second order. We will do this in turn in the next two sections.

Before concluding the present section, though, let us perform a consistency check on the Lagrangian (\ref{Lz}). Namely, let us verify that it does have as a solution of its equations of motion the same classical configuration as (\ref{clSol}), (\ref{derVarPhi}). Turning off the $x^{\mu}$-dependence, which only arises at the level of fluctuations, we have from the square root in (\ref{Lz}):
\be
L_c = \sqrt{3(1+z^2) \!\left[\theta_z^2+\sin^2 \!\theta \varphi_z^2\right]+e^{4 \rho_0}\beta\,z^2} \,\, ,
\ee
where we have also divided by an overall factor of $2$. The variational equations for $\theta (z)$ and $\varphi (z)$ are then:
\bea
\frac{d}{dz}\left\{\frac{(1+z^2)\sin^2 \!\theta\,\varphi_z}{L_c}\right\} &=& 0 \, ,\label{phivar}\\
\frac{d}{dz}\left\{\frac{(1+z^2)\,\theta_z}{L_c}\right\} - \frac{(1+z^2) \sin(2\theta) (\varphi_z)^2}{L_c} &=& 0 \, . \label{thetavar}
\eea
Equation (\ref{thetavar}) is solved by $\theta = \pi/2$. As a result, (\ref{phivar}) simplifies to:
\be\label{phivar2}
\frac{d}{dz}\left\{\frac{(1+z^2)\,\varphi_z}{\sqrt{3\,(1+z^2)\,\varphi_z^2 + e^{4\rho_0} \beta \,z^2}}\right\}=\,0 \,\, .
\ee
The latter equation is solved by:
\be\label{classical1}
\varphi_z(z)=e^{2\rho_0}\sqrt{\frac{\beta}{3}}\frac{1}{\sqrt{1+z^2}} \,\, ,
\ee
which is the same as (\ref{derVarPhi}). For completeness, let us also note that (\ref{classical1}) implies
\be\label{classicalphi}
\varphi (z) = e^{2\rho_0}\sqrt{\frac{\beta}{3}} \,\,{\rm arcsinh} (z) \, ,
\ee
where we have set the integration constant to zero in order to have $\varphi = 0$ at $z=0$ (in other words, in order to have the ${\rm D}7$ and $\overline{{\rm D}7}$ branches merge at $z=0$). Finally, note that the profile (\ref{classicalphi}) agrees with the one determined by (4.12) of \cite{LA}. Namely, in our approximations the constant $B$ there becomes $B = \frac{\beta}{3}$; in addition, the plus sign there should be taken for $z>0$ while the minus sign for $z<0$.

\section{Spectrum of $\theta$ bosons}
\setcounter{equation}{0}

In this section we begin investigating the spectrum of scalar mesons in our model. The latter arise from the fluctuations of the embedding functions $\theta$ and $\varphi$ around the U-shaped D7-$\overline{{\rm D}7}$ embedding, that realizes chiral symmetry breaking. First, we address the spectrum of $\theta$ bosons. In the next subsection we derive the relevant equation of motion and show that the resulting mass spectrum differs from the one for vector mesons, computed in \cite{ASW}, only by an order $\beta$ correction. Finally, in Subsection \ref{ThetaMass} we find explicitly the ${\cal O} (\beta)$ mass difference between the spectra of $\theta$ bosons and vector mesons.

\subsection{Equation of motion}

Let us now turn to studying the spectrum of fluctuations of the field $\theta(z,x^{\mu})$ around the configuration (\ref{clSol}). To do that we have to substitute (\ref{tpFluct}) in (\ref{Lz}) and expand to second order in the fluctuations. Actually, as already pointed out, we can obtain the relevant $\delta \theta$ Lagrangian by first setting to zero the $\varphi$ fluctuations and then expanding. We also need to substitute (\ref{derVarPhi}). The result is:
\be \label{Lz2}
{\cal L} = const \frac{1}{\sqrt{1+z^2}}\left\{12(1+z^2)(\pd_z \delta\theta)^2+e^{4\rho_0}\beta\left[ c\,(1+z^2)\!\left( \pd_{\mu} \delta\theta \right)^2-4(\delta \theta)^2 \right] \right\}.
\ee
Using that $\pd_{\mu} \pd^{\mu} \delta \theta = m_{\theta}^2 \delta \theta$, we then find the following field equation:
\be \label{eq1}
-(\delta \theta)''(z)-\frac{z}{1+z^2}(\delta \theta)'(z)-\frac{\beta\,e^{4 \rho_0}}{3(1+z^2)}\delta \theta(z)=M^2 \delta \theta (z) \, ,
\ee
where $'\equiv \pd_z$ and $M^2 = \frac{m_{\theta}^2 c \beta e^{4\rho_0}}{12}$; also, we have suppressed the $x^{\mu}$ argument for brevity. To bring (\ref{eq1}) to Schrodinger form, we perform the field redefinition:
\be
\delta \theta(z)= \frac{\Theta(z)}{(1+z^2)^{1/4}} \,\, .
\ee
Then the equation for the mass spectrum becomes:\footnote{Let us make an important remark. Clearly, the potential term in the Schrodinger equation (\ref{eqtheta}) changes sign as $z$ is varied. However, the presence of a negative potential region does not by itself indicate an instability. The configuration we are expanding around would be unstable if the spectrum of solutions of (\ref{eqtheta}) with appropriate boundary conditions were to have a negative mass-squared mode. In the rest of this Section, we will show that this is not the case.}
\be\label{eqtheta}
-\Theta''(z)-\left[\frac{z^2-2}{4(1+z^2)^2}-V_1(z)\right]\Theta(z)=M^2\Theta (z) \, ,
\ee
where
\be \label{V1z}
V_1(z)=- \frac{\beta\,e^{4\rho_0}}{3(1+z^2)} \, .
\ee

Before we turn to solving (\ref{eqtheta}), let us make a few comments on its form. First of all, notice that, except for the $V_1 (z)$ correction term, this equation is exactly the same as (4.4) of \cite{ASW}.\footnote{In that regard, note that the parameter $\lambda$ in (4.5) of \cite{ASW} can be removed by introducing a new variable $\hat{z} = z/\sqrt{\lambda}$ in (4.4) there. In addition, using that $c_0 \equiv c \sqrt{\beta}$ and the expression for $\lambda$ in (4.6) of \cite{ASW}, one can see that the new variable $\hat{z}$ is precisely the same as the variable $z$ in (\ref{ChVar}) here.} The latter determines the spectrum of vector and axial-vector mesons in this model. Therefore, the presence of the term (\ref{V1z}) here implies an ${\cal O} (\beta)$ splitting between the masses of $\theta$ bosons and vector mesons. Actually, at first sight one might think that such a conclusion is premature, as in \cite{ASW} we were working at small $\beta$ as well and keeping only the leading contributions. So it is legitimate to suspect that the potential in (4.4) of \cite{ASW} might have an order $\beta$ correction that we have just not written down. However, we will show now that any such correction would be the same for both the vector bosons and $\theta$ mesons. Hence the correction (\ref{V1z}) does, in fact, give rise precisely to the mass splitting between the $\theta$ and vector meson spectra.

To understand that, it is most instructive to derive the full field equation for $\delta \theta$, without using any of the approximations that led to the simplified walking metric (\ref{leadMetric}). Working with the full metric (\ref{BM}) and introducing the same functions $a(\rho)$ and $b(\rho)$ as in (2.21) of \cite{ASW}, namely
\bea \label{ab}
a(\rho) &=& \frac{1}{A^3} \left[ f(\rho) + g(\rho) \varphi_{\rho}^2 \right]^{1/2} \nn \\
b(\rho) &=& \frac{1}{A^2} \left[ f(\rho) + g(\rho) \varphi_{\rho}^2 \right]^{1/2} g^{\rho \rho}
\eea
with
\be
f(\rho) = \frac{A^8 c^4}{256} P_1^2 (\rho) P_1'^2 (\rho) \coth^2 (2 \rho) \quad \, , \quad \, g(\rho) = \frac{A^8 c^4}{2\times 256} P_1^3 (\rho) P_1' (\rho) \coth(2 \rho)
\ee
\be
\hspace*{-0.1cm}{\rm and} \quad \, g^{\rho \rho} = \frac{A^7 c^3 P_1^2 P_1' \coth^2 (2 \rho)}{128 \left( f + g \varphi_{\rho}^2 \right)} \,\, ,
\ee
we find after a lengthy calculation:
\be\label{rhoeq}
-\partial_\rho\,[\, b(\rho)\, \partial_\rho \,\delta \theta(\rho)\,]-b(\rho)\coth^2(2\rho)[\varphi_{\rho}^{cl}(\rho)]^2 \delta \theta(\rho)=m_{\theta}^2\,a(\rho)\,\delta \theta(\rho) \, .
\ee

The above equation of motion differs from the (axial-)vector meson one, eq. (2.26) of \cite{ASW}, only by the presence of the $[\varphi_{\rho}^{cl}]^2$ term. Although it is by no means easier to try to solve (\ref{rhoeq}) instead of (\ref{eqtheta}), it is instructive to see how (\ref{rhoeq}) reduces to (\ref{eqtheta}) in the approximations that give the walking metric (\ref{leadMetric}). Namely, to leading order in small $\beta$, we find that the functions (\ref{ab}) reduce to:\footnote{To obtain (\ref{abrho}), one also needs to use the relation $c \sqrt{\beta} = \frac{\sqrt{3}}{16}$, valid to leading order in small $\beta$, that was derived in \cite{ASW}.}
\bea \label{abrho}
a(\rho (z)) &=& \frac{3^{1/4} c^{5/4} \beta^{3/4} e^{4\rho_0}}{24} \,\frac{\left( 1+z^2 \right)^{3/2}}{z}\nn \\
b(\rho (z)) &=& \frac{3^{1/4} c^{1/4} \beta^{-1/4}}{8} \,\frac{z}{\sqrt{1+z^2}} \,\, ,
\eea
where we have used (\ref{ChVar}). It is then easy to see that (\ref{rhoeq}) becomes exactly (\ref{eq1}), after the change of variable from $\rho$ to $z$. The important point is that the term $-\frac{\beta e^{4\rho_0}}{3 (1+z^2)} \delta \theta (z)$ in (\ref{eq1}), that gives rise to $V_1 (z)$, comes entirely from (and in fact, to leading order, is the only contribution of) the $[\varphi_{\rho}^{cl}]^2$ term in (\ref{rhoeq}), which is precisely the extra term not present for vector bosons. Hence, whatever other ${\cal O} (\beta)$ contributions might arise from the remaining terms in (\ref{rhoeq}), when subleading orders in $a(z)$ and $b(z)$ are taken into account, they will necessarily be the same for both the vector and $\theta$ mesons. Thus the leading order mass  difference between the two spectra is determined precisely by $V_1 (z)$, while any subleading correction to it will come from higher order terms in the small $\beta$ expansion of the $[\varphi_{\rho}^{cl}]^2$ term.

\subsection{Mass spectrum} \label{ThetaMass}

Let us now consider the Schrodinger equation (\ref{eqtheta}) in more detail. Clearly, at leading order in small $\beta$, the $V_1 (z)$ term in the potential can be neglected and so one has exactly the same spectrum as for the vector mesons, studied in \cite{ASW}. To evaluate the mass split between the two spectra due to $V_1 (z)$, we will use perturbation theory. Note also that, since we are studying the perturbative spectrum, we will impose the boundary conditions that at $z=\pm z_{\Lambda}$ the wave function vanish.\footnote{Here, as in \cite{ASW}, we have introduced a physical cut-off $z_{\Lambda}$ at the upper end of the walking region, so that $z\in (-z_{\Lambda}, z_{\Lambda})$. The earlier work \cite{LA} discussed holographic renormalization for the background of interest. However, in \cite{ASW} we pointed out that above the scale $z_{\Lambda}$ the light spectrum of the model should change in order to reflect the degrees of freedom appropriate for extended technicolor. Hence, on physical grounds, our current model should not be viewed as valid to arbitrarily high energies. The proper UV completion is still an open problem; see \cite{CGNPR}, though, for important recent progress in that direction.} Let us denote by $\Theta_n^0$ the zeroth order wave function of the $n^{{\rm th}}$ state, i.e. with $V_1(z)$ neglected. Then, according to standard rules, the correction to the mass-squared $M_n^2$ due to the perturbation $V_1(z)$ is given by:
\be \label{Mn2}
\delta \,[M_n^2] =\frac{1}{N^2}\int_{-z_\Lambda}^{z_\Lambda} [\Theta_n^0 (z)]^2 \,V_1(z)\,dz \, ,
\ee
where the normalization factor is
\be \label{norm}
N^2=\int_{-z_\Lambda}^{z_\Lambda} [\Theta^0_n (z)]^2\,dz \, .
\ee

From the analysis of Section 4 in \cite{ASW}, we know the approximate analytical form of $\Theta_n^0 (z)$. More precisely, one can divide the interval $(0, z_{\Lambda})$ into two regions, in each of which the field equation simplifies and can be solved analytically. The matching of the two then gives the quantization condition $M_n = r_n / z_{\Lambda}$, where $r_n$ is the $n^{{\rm th}}$ root of the Bessel function $J_0$. Up to overall constants, the wave function has the following form in each of the two regions:
\be \label{Tworegions}
\Theta_n^0 (z) = \begin{cases} \Theta_n^{(s)} (z) = (1 + z^2)^{1/4}\,,& \mbox{\,\,\,\,\,for small $z$, \,i.e. $\!z \in (0, z_{\bullet})$ \,,}\\ \Theta_n^{(l)} (z) = \sqrt{z} \,J_0 (M_n z)\,, & \mbox{\,\,\,\,\,for large $z$, \,i.e. $\!z \in (z_{\bullet} , z_{\Lambda})$ \,,}
\end{cases}
\ee
where $1<\!\!< z_{\bullet} <\!\!< z_{\Lambda}$.\footnote{For simplicity, we took in (\ref{Tworegions}) the form of the solution, which is symmetric under $z \rightarrow -z$. The antisymmetric solution, that now gives pseudoscalar mesons, is a bit more involved technically as can be seen from the axial-vector considerations in \cite{ASW}. However, the final outcome is the same, except for substituting $r_n$ with the quantity $\mu_n$ defined in (4.16) of \cite{ASW}.} Actually, it is easy to realize that one can approximate the solution on the entire $0<z<z_\Lambda$ interval (except for a $\Delta z\simeq M^2$ neighborhood of $z=\sqrt{2}$) by the following function:
\be\label{interpolate}
\hat{\Theta}_n^0(z)= (1 + z^2)^{1/4}\,J_0(M_n z) \,\, .
\ee
Clearly, $\hat{\Theta}$ reduces to the first line of (\ref{Tworegions}) at small $z$, while it acquires the form on the second line of (\ref{Tworegions}) at large $z$. To see by how much this approximation fails to be an exact solution, let us substitute $\Theta = \hat{\Theta}$ in (\ref{eqtheta}), with $V_1$ omitted. We obtain:
\be\label{eqtheta0}
-\hat{\Theta}_n^0{}''(z)-\left[\frac{z^2-2}{4(1+z^2)^2}+M_n^2\right]\hat{\Theta}_n^0(z)=-\frac{M_n\,J_1(M_n z)}{z(1+z^2)^{3/4}} \,\, ,
\ee
where the nonvanishing of the right hand side is exactly the measure of the deviation of $\hat{\Theta}$ from being an exact solution. It is easy to see that, aside from a small neighborhood of a finite number of points, this right hand side is of ${\cal O}(z_\Lambda^{-2})$ compared to the individual terms on the left hand side on the whole interval $0<z<z_\Lambda$.

In evaluating the integral in (\ref{norm}), it is easy to realize that the leading contribution in the small $\beta$ expansion will come entirely from the large $z$ region. So, to leading order, we have:
\be
N^2=\int_{-z_\Lambda}^{z_\Lambda} [\Theta^0_n (z)]^2\,dz \simeq\int_{-z_\Lambda}^{z_\Lambda} z[J_0(M_n\,z)]^2\,dz= z_{\Lambda}^2 \, [J_1 (M_n z_{\Lambda})]^2 \, .
\ee
On the other hand, the integral in (\ref{Mn2}) can be estimated by using the approximate solution $\hat{\Theta}_n^0(z)$. To leading order in $z_\Lambda$, one obtains:
\be \label{SmallZ}
\int_{- z_{\Lambda}}^{z_{\Lambda}} [\hat{\Theta}_n^0 (z)]^2 V_1(z) dz = -\frac{2}{3} \beta e^{4 \rho_0} \!\left[\,{\rm arcsinh} (z_{\Lambda}) -\frac{1}{4}r_n^2\,\,{}_3F_4 (1,1,\frac{3}{2};2,2,2,2;-r_n^2)+ {\cal O} (z_\Lambda^{-2})\right]\!,
\ee
where ${}_3F_4 (1,1,\frac{3}{2};2,2,2,2;-r_n^2)$ is the hypergeometric function. The details of this computation are somewhat subtle and we have relegated them to Appendix B. Combining the above results, we find to leading order in $z_\Lambda$:
\be
\delta \,[M_n^2] =  -\frac{2}{3} \,\frac{\beta \,e^{4 \rho_0}}{r_n{}^2J_1{}^2(r_n)}\,M_n^2 \left[\,{\rm arcsinh} (z_{\Lambda}) -\frac{1}{4}r_n^2\,\, {}_3F_4 (1,1,\frac{3}{2};2,2,2,2;-r_n^2) +{\cal O} (z_\Lambda^{-2})\right],
\ee
where we have used $z_{\Lambda} = r_n / M_n$. Now, recalling that $M^2 = \frac{m_{\theta}^2 c \beta e^{4\rho_0}}{12}$ and also using that for large $z_{\Lambda}$ the function ${\rm arcsinh} (z_{\Lambda}) = \ln ( z_{\Lambda} + \sqrt{1+ z^2_{\Lambda}} )$ simplifies to $\ln (2 z_{\Lambda})$, we can write the correction to the $\theta$ boson mass spectrum, i.e. $\delta (m_n^{\theta})^2 = (m^{\theta}_n)^2-(m_n^V)^2$, as:
\be\label{shift}
\delta (m ^\theta_n)^2=-\,m^2_n\,\frac{2\,\beta\,e^{4\rho_0}}{3\,[r_n J_1(r_n)]^2} \left[ \,\log(2 z_\Lambda) -\frac{1}{4}r_n^2\,\, {}_3F_4 (1,1,\frac{3}{2};2,2,2,2;-r_n^2) + {\cal O}(z_\Lambda^{-2})\right] ,
\ee
where the unperturbed (i.e., with $V_1 (z)$ neglected) mass of the $n^{{\rm th}}$ state is \cite{ASW}:
\be \label{masssq}
m_n^2= \frac{12}{c_0}\beta^{-1/2} e^{-4\rho_\Lambda} \,r_n^2 \,M_{KK}^2
\ee
with $c_0 = \frac{\sqrt{3}}{16}$ to leading order.

As can be seen from (\ref{shift}), the mass split between the $n^{\rm th}$ vector and $\theta$ boson states is suppressed by a power of $\beta$ compared to the leading contribution. It would be very interesting to understand the origin of this split, or rather the reason for its suppression. We leave this for future work.

\section{Spectrum of $\varphi$ bosons}
\setcounter{equation}{0}

In this section we will study the fluctuations of the field $\varphi (z,x^{\mu})$ around the classical solution (\ref{clSol}), (\ref{derVarPhi}). As before, to obtain the Lagrangian of interest we have to substitute (\ref{tpFluct}) in (\ref{Lz}) and expand to second order in $\delta \varphi (z,x^{\mu})$ and $\delta \theta (z, x^{\mu})$. However, as already pointed out, to quadratic order there is no mixing between the $\theta$ and $\varphi$ fluctuations. So we can safely set $\theta = \pi /2$ from the start. Hence, the Lagrangian (\ref{Lz}) acquires the form:
\be\label{Lz1}
{\cal L} = const \sqrt{12(1+z^2) \left(\varphi'_{cl} + \delta \varphi'\right)^2+e^{4\rho_0}\beta\,z^2\left[ 4+c\, (\pd_{\mu} \delta \varphi \right)^2]} \,\, ,
\ee
where again $' = \pd_z$. Expanding to second order and substituting (\ref{derVarPhi}), we have:
\be\label{Leff1}
{\cal L} = \frac{3 e^{-2\rho_0}}{\sqrt{\beta}}\frac{z^2}{\sqrt{1+z^2}} \, (\delta \varphi')^2+2\sqrt{3}\, \delta \varphi' + \frac{e^{2\rho_0} c \, \sqrt{\beta}}{4} \frac{z^2}{\sqrt{1+z^2}} \, ( \pd_{\mu} \delta \varphi )^2 \, ,
\ee
where for convenience we have dropped the overall constant in front of the square root in (\ref{Lz1}); also, we have omitted the additive term $2 e^{2\rho_0} \sqrt{\beta} \sqrt{1+z^2}$\,, since clearly it drops out of the field equation. Now, varying (\ref{Leff1}) and using $\pd_{\mu} \pd^{\mu} \delta \varphi = m_{\varphi}^2 \delta \varphi$, we obtain the following equation of motion:
\be\label{eq2}
-\,\delta \varphi''(z)-\frac{2+z^2}{z(1+z^2)} \,\delta \varphi'(z)=M_\varphi^2 \,\delta \varphi(z) \, ,
\ee
where $M^2_{\varphi} = m^2_{\varphi} c \beta e^{4 \rho_0} / 12$ and we have suppressed the argument $x^{\mu}$ for brevity.

In line with the previous section (as well as our considerations in \cite{ASW}), it would seem that the strategy to study (\ref{eq2}) invloves first transforming to Schrodinger form. For technical reasons, however, we will use another approach. Despite that, let us briefly discuss for completeness the Schrodinger form of equation (\ref{eq2}). It is achieved via the field redefinition:
\be
\delta \varphi (z) = \frac{(1+z^2)^{1/4}}{z} \,\Phi (z) \, ,
\ee
which leads to:
\be \label{VpSchr}
- \Phi'' (z) - \frac{6+z^2}{4(1+z^2)^2} \,\Phi (z) = M_{\varphi}^2 \,\Phi (z) \, .
\ee
This is rather similar, although not quite the same as the equation for the vector mesons. Recall that the latter is
\be \label{VSchr}
- \Psi'' (z) - \frac{z^2 - 2}{4(1+z^2)^2} \,\Psi (z) = M_V^2 \,\Psi (z) \, ,
\ee
as explained in the previous section. There we found that the difference between the $\theta$ and vector meson cases was given by the order $\beta$ correction to the potential $V_1 (z)$ in (\ref{V1z}). On the other hand, (\ref{VpSchr}) and (\ref{VSchr}) differ by a potential difference $\Delta V = \frac{2}{(1+z^2)^2}$. At large $z$, one has $\Delta V \sim \frac{2}{z^4}$, which is negligible compared to the large $z$ behaviour of each of the two potentials in (\ref{VpSchr}) and (\ref{VSchr}), that is $\frac{1}{4 z^2}$. However, at small $z$ the difference is of the same order as the potentials themselves. So we cannot view $\Delta V$ as a small perturbation compared to the vector meson case. Nevertheless, one may expect that the mass difference between the two cases is small due to the same large $z$ asymptotics they exhibit.

Since we cannot solve exactly either of (\ref{VpSchr}) and (\ref{VSchr}), it turns out that the most efficient way to compare their mass spectra is to take a step back and to not go to Schrodinger form. That is, for the $\varphi$ bosons we will consider (\ref{eq2}); for future use let us also introduce the notation:
\be
H_{\varphi} = - \frac{d^2}{dz^2} - \frac{2+z^2}{z(1+z^2)} \frac{d}{dz}
\ee
for the differential operator on its left-hand side. Clearly, $H_{\varphi}$ can be viewed as a Hamiltonian operator with eigenfunctions $\delta \varphi (z)$ and eigenvalues $M^2_{\varphi}$\,. The analogue of (\ref{eq2}) for the vector bosons is
\be \label{eq3}
- \psi'' (z) - \frac{z}{1+z^2} \,\psi' (z) = M_V^2 \,\psi(z) \, ,
\ee
where
\be \label{psired}
\psi (z) = \frac{\Psi (z)}{(1+z^2)^{1/4}}
\ee
with the same $\Psi (z)$ as in (\ref{VSchr}). Note that in \cite{ASW} we did not write down equation (\ref{eq3}) explicitly. Instead, we obtained directly (\ref{VSchr}), because we combined the change of variables from $\rho$ to $z$ with the redefinition (\ref{psired}). Indeed, the factor $\frac{1}{[a(\rho) b(\rho)]^{1/4}}$ in (4.3) of \cite{ASW} accounts precisely for the factor $\frac{1}{(1+z^2)^{1/4}}$ in (\ref{psired}) here. As a last preparation, let us also introduce the notation:
\be
H_V = - \frac{d^2}{dz^2} - \frac{z}{1+z^2} \frac{d}{dz}
\ee
for the differential operator on the left-hand side of (\ref{eq3}).

Now, our goal will be to compare the Hamiltonians $H_{\varphi}$ and $H_V$. Since, as already pointed out above, the mass difference is expected to be small, we can view $\Delta H = H_{\varphi} - H_V$ as a small perturbation. Of course, once we obtain the answer, it will be easy to verify that this assumption is satisfied. Before proceeding further, it is very useful to notice the following relation:
\be \label{3Hs}
H_{\varphi} - H_V = 2 (H_{\varphi} - H_0) \, ,
\ee
where the operator $H_0$ is
\be
H_0 = - \frac{d^2}{dz^2} - \frac{1}{z} \frac{d}{dz} \,\, .
\ee
The differential equation $H_0 \chi (z) = M^2 \chi(z)$, namely
\be \label{chieq}
- \chi'' (z) - \frac{1}{z} \chi' (z) = M^2 \chi (z) \, ,
\ee
is in fact precisely of the form that both (\ref{eq2}) and (\ref{eq3}) acquire at large $z$. Unlike those equations, though, (\ref{chieq}) can be solved easily. The regular solutions, that vanish at $z=\pm z_{\Lambda}$, are given by:
\be
\chi_n (z) = J_0 \!\left( \frac{r_n z}{z_{\Lambda}} \right) ,
\ee
where $r_n$ is again the $n^{{\rm th}}$ root of the Bessel function $J_0$. The key use of (\ref{3Hs}) is therefore that, since we know the eigenfunctions of $H_0$, we can compute the mass shift, due to $\Delta H$, by using standard first order perturbation theory:
\be \label{DeltaM2}
\Delta M^2_n = \frac{2}{\langle \chi_n | \chi_n \rangle} \,\langle \chi_n | \Delta H_0 | \chi_n \rangle \, ,
\ee
where $\Delta H_0 \equiv H_{\varphi} - H_0$. And, of course, since (\ref{chieq}) is the same as the large $z$ asymptotic form of (\ref{eq2}), the perturbation $\Delta H_0$ is expected to be small just as $\Delta H = H_{\varphi} - H_V$ is. Before turning to the computation of $\Delta M_n^2$ according to (\ref{DeltaM2}), we also need to note the following. Hermiticity of the Hamiltonian $H_0$ requires the use of the measure $|z| dz$. Indeed, one can easily show that
\be
\int_{-z_{\Lambda}}^{z_{\Lambda}} \chi_m (z) \chi_n (z) \,|z| \,dz \,\sim \,\delta_{mn}
\ee
and
\be
\int_{-z_{\Lambda}}^{z_{\Lambda}} \chi_m (z) \,H_0 \,\chi_n (z) \,|z| \,dz \,= \int_{-z_{\Lambda}}^{z_{\Lambda}} \chi_n (z) \,H_0 \,\chi_m (z) \,|z| \,dz  \,\sim \,\delta_{mn} \,\, .
\ee
\newline
\indent
Now we are finally ready to compute $\Delta M_n^2 = [M_n^{\varphi}]^2 - [M_n^V]^2$. Substituting the explicit form of $\chi_n (z)$ in (\ref{DeltaM2}), we find:
\be
\left[M_n^\varphi\right]^2-\left[M_n^V\right]^2=-\frac{4 }{N_n^2}\int_0^{z_\Lambda} z\,J_0(M_n\,z)\frac{1}{z(1+z^2)}\frac{d}{dz}J_0(M_n\,z)\,dz \, ,
\ee
where
\be
N_n^2=2\int_0^{z_\Lambda}z \,[J_0(M_n\,z)]^2 dz=z_\Lambda^2[J_1(r_n )]^2 \, .
\ee
On the last line we have used $J_0(M_n z_\Lambda)=0$. After substituting $x= M_n z$\,, we obtain:
\bea \label{Mnvphi}
\left[M_n^\varphi\right]^2-\left[M_n^V\right]^2&=&-\frac{2}{N_n^2}\int_0^{r_n}\frac{M_n^2}{M_n^2+x^2}\frac{d}{dx}[J_0(x)]^2dx\nn\\&=&\frac{2}{N_n^2}M_n^2[\log(1/M_n)+c_n+O(M_n^2\,\log(M_n))] \, ,
\eea
where the constants $c_n$ are determined by:
\be
c_n=\frac{1}{r_n^2}+\log(r_n^2)-\frac{1}{2}-2\int_0^{r_n}\frac{1}{x^3}\left[J_0^2(x)+\frac{x^2}{2}-1\,\right] dx \, .
\ee
Equivalently, we can rewrite (\ref{Mnvphi}) as:
\be\label{phidev}
\Delta (m_n^\varphi)^2 = (m_n^\varphi)^2 - (m_n^V)^2 = m_n^2 \frac{2}{z_\Lambda^2[J_1(r_n )]^2}[\log(z_\Lambda/r_n)+c_n+O(z_\Lambda^{-2}\,\log(z_\Lambda)\,)] \, ,
\ee
where again $m_n^2$ is given by (\ref{masssq}). For example, for the lowest lying state $c_1\simeq  0.619$.
Thus, for $n=1$ we have:
\be
(m_1^{\varphi})^2-(m_1^V)^2=m_1^2 \,\frac{3.86}{z_\Lambda^2}\,[\log(z_\Lambda)+0.887+O(z_\Lambda^{-2}\,\log(z_\Lambda)\,)] \, .
\ee

It is interesting to compare the mass differences between the two kinds of scalar bosons, $\theta$ and $\varphi$, and the vector mesons. First, from (\ref{shift}) and (\ref{phidev}) one can see that $\Delta (m^\theta_n)^2$ and $\Delta (m_n^{\varphi})^2$ have opposite signs. More precisely, the $\theta$ bosons are slightly lighter, while the $\varphi$ mesons are slightly heavier than the vector bosons. Despite that, the magnitudes of $\Delta (m^\theta_n)^2$ and $\Delta (m_n^{\varphi})^2$ are almost comparable. This is because $\beta \sim z_{\Lambda}^{-2}$, which follows from $z_{\Lambda} \equiv z(\rho_{\Lambda}) \approx e^{2 \rho_{\Lambda}}$ as well as the relation $e^{2 \rho_{\Lambda}} \sim \beta^{-1/2}$ that was elaborated on in \cite{ASW}. However, to be more precise, we should write $e^{4 \rho_{\Lambda}} = \sigma \beta^{-1}$, where $\sigma$ is a number of order $10^{-1}$ or $10^{-2}$ \cite{ASW}. Therefore, the ratio $\Delta (m^\theta_n)^2 / \Delta (m_n^{\varphi})^2 \sim \sigma <1$ and thus the $\varphi$ boson shift is somewhat greater than the $\theta$ boson one.

\section{Discussion}

We investigated the spectrum of scalar mesons in the gravity dual of a walking technicolor model, obtained by embedding D7-$\overline{{\rm D}7}$ techniflavor probes \cite{LA,ASW} in the technicolor background of \cite{NPP}. We showed that all mass-squareds are positive and thus there is no tachyonic mode with respect to fluctuations of the probe branes embedding. Furthermore, we found that both types of scalar mesons, arising from those fluctuations, have spectra that are nearly degenerate with the vector meson spectrum, studied in \cite{ASW}.\footnote{Interestingly, this is in line (albeit, more pronounced in our case) with the expectations, based on the use of the Schwinger-Dyson and Bethe-Salpeter equations, that in IR-fixed-point walking theories the masses of scalar and vector mesons at a given level are numerically close to each other \cite{KSh}.} Namely, the mass differences are of higher order, in an expansion in the small parameter $\beta$, compared to the leading contributions to the masses. It would be very interesting to understand the underlying reason for this near-degeneracy. Likely, it will mostly turn out to be due to a small amount of supersymmetry breaking, as the massive vector and scalar states should belong to the same long ${\cal N} = 1$ supermultiplet.\footnote{Such an expectation is supported by the results of \cite{NPR}, where the flavor probe embedding is supersymmetric and the spectra of vector and scalar mesons are exactly degenerate. We thank C. Nunez for pointing this out.} However, the difference between the mass shifts of the two kinds of scalars may indicate a role for an additional factor.
We leave
investigation of this question for future work.

One should keep in mind that the scalar spectrum studied here does not address the question of whether our walking technicolor model has a dilaton or not. The latter is also a scalar field. However, it is a mode that, for us, should arise predominantly from fluctuations of the technicolor background (see the discussion in the introduction). In particular, its distinguishing feature in the effective Lagrangian is a certain coupling to the square of the technicolor field strength.\footnote{Note that \cite{ENP} found evidence for a light scalar in the spectrum of fluctuations of the color background of \cite{NPP}. However, at present, it is unclear whether this state can be identified with the technidilaton.} The phenomenological importance of this field is great, since it is a (potentially) light scalar boson rather similar to the Higgs. To differentiate between the two, one would have to compute their couplings to other fields; for such a study see \cite{GGS}. There is a lot of literature for \cite{Yam,BLL,Sannino,AB} or against \cite{NSY,against,KLP} the existence of a dilaton in walking technicolor.\footnote{It is also worth mentioning the pioneering work of \cite{GW} on Goldstone bosons for broken conformal invariance, albeit in the case of {\it weakly-coupled} gauge theories.} The arguments for are inspired by the approximate conformal symmetry of the walking region, whereas the arguments against rely on the fact that the theory is never exactly conformal. We hope to investigate the issue of the presence of a technicolor dilaton in our model in a future publication.

Other important open problems are the computation of the $T$ and $U$ electroweak observables, as well as the anomalous dimension of the technifermion condensate. Finally, there are various interesting possibilities for extending our model. An important issue, equally relevant for all other D-brane constructions of technicolor models \cite{HolTech}, is to go beyond the flavor probe approximation. This would enable one to study gravity duals of models with comparable numbers of colors and flavors, like the walking technicolor models that rely on the existence of a Banks-Zaks fixed point for the gauge coupling in the infrared region \cite{BZ}. Another interesting direction, relevant for understanding the generation of quark and lepton masses, is to try to build a gravity dual of extended technicolor \cite{extTC}. In that regard, the recent work \cite{CGNPR} might provide valuable insights.

\section*{Acknowledgements}

We are grateful to T. Appelquist, P. Argyres, S. Das, Z. Komargodski, M. Kruczenski, C. Nunez, M. Piai, F. Sannino, G. Semenoff, A. Shapere and R. Shrock for useful discussions. In addition, L.A. thanks the October, 2011, SPOCK meeting in Cincinnati and the 2011 Simons Workshop in Stony Brook for hospitality during various stages of this work. Research at Perimeter Institute is supported by the Government of Canada through Industry Canada and by the Province of Ontario through the Ministry of Research \& Innovation. L.A. is also supported in part by funding from NSERC. The research of P.S. and R.W. is supported by DOE grant FG02-84-ER40153.

\appendix

\section{Determinant of induced metric}
\setcounter{equation}{0}

Here we compute the determinant of the fluctuated metric, induced on the ${\rm D}7$-$\overline{{\rm D}7}$ worldvolume.

Performing the change of variable (\ref{ChVar}), the metric (\ref{leadMetric}) acquires the form:
\be \label{zmetric}
ds^2_{{\rm walk}} = A \left[ \eta_{\mu \nu} dx^{\mu} dx^{\nu} + \frac{cP_1'}{8} \left( 4 \left( \pd_z \rho \right)^2 dz^2 + \left( \omega_3 + \tilde{\omega}_3 \right)^2 \right) + \frac{c}{4} \left( d \Omega_2^2 + d \tilde{\Omega}_2^2 \right) \right] ,
\ee
where recall that $\omega_3 = \cos \theta \,d \varphi$ and $d \Omega_2^2 = d \theta^2 + \sin^2 \!\theta \,d \varphi^2$; also, now we have:
\be \label{P1p}
P_1' = \frac{2}{3} \beta e^{4\rho_0} (z^2 + 1) \, .
\ee
Substituting (\ref{d1}) in (\ref{zmetric}), we find the following non-vanishing metric components:
\bea \label{8dmetric}
g_{\mu \nu} &=& A \left[ \eta_{\mu \nu} + \frac{c P_1'}{8} (\cos^2 \!\theta) \varphi_{\mu} \varphi_{\nu} + \frac{c}{4} \theta_{\mu} \theta_{\nu} + \frac{c}{4} (\sin^2 \!\theta) \varphi_{\mu} \varphi_{\nu} \right] \nn \\
g_{\mu z} &=& A \left[ \frac{cP_1'}{8} (\cos^2 \!\theta) \varphi_{\mu} \varphi_z + \frac{c}{4} \theta_{\mu} \theta_z + \frac{c}{4} (\sin^2 \!\theta) \varphi_{\mu} \varphi_z \right] \nn \\
g_{z z} &=& A \left[ \frac{cP_1'}{8} \left( 4 \left( \pd_z \rho \right)^2 + \cos^2\!\theta \,\varphi_{z}^2 \right) + \frac{c}{4} (\theta_z^2 + \sin^2 \!\theta \,\varphi_z^2) \right] \nn \\
g_{z \tilde{\omega}_3} &=& A \frac{cP_1'}{8} \cos \!\theta \,\varphi_z  \qquad , \qquad g_{\mu \tilde{\omega}_3} \,\,\, = \,\,\, A \frac{cP_1'}{8} \cos \!\theta \,\varphi_{\mu} \nn \\
g_{\tilde{\omega}_3 \tilde{\omega}_3} &=& A \frac{cP_1'}{8} \qquad , \qquad g_{\tilde{\omega}_1 \tilde{\omega}_1} = A \frac{c}{4} \qquad , \qquad g_{\tilde{\omega}_2 \tilde{\omega}_2} = A \frac{c}{4} \qquad .
\eea
Hence, we have that $\det g_{8d} = A^2 \left( \frac{c}{4} \right)^2 \det g_{6d}$, where $g_{6d}$ is the 6d metric along the $\mu$, $\rho$ and $\tilde{\omega}_3$ directions.

Now, we can use that
\be
\det g_{6d} = g_{\tilde{\omega}_3 \tilde{\omega}_3} \det G_{5d} \, ,
\ee
where $G_{5d}$ is a 5d metric with the following components:\footnote{This is, clearly, a special case of the general relation \,$\det \!\left( \begin{array}{cc} A & B \\ C & D \end{array} \right) \!= \det A \times \,\det \left( D - C A^{-1} B \right)$.}
\be \label{GMN}
G_{MN} = g_{MN} - \frac{g_{M \tilde{\omega}_3 } g_{N \tilde{\omega}_3}}{g_{\tilde{\omega}_3 \tilde{\omega}_3}} \qquad {\rm with} \,\,\, M,N \,\,\,{\rm running \,\,\,over} \,\,\,\mu, z \,\,\, .
\ee
Hence
\be
\det g_{8d} = A^3 \left( \frac{c}{4} \right)^2 \frac{cP_1'}{8} \det G_{5d}
\ee
where, more explicitly, the components of $G_{5d}$ are:
\bea
G_{\mu \nu} &=& g_{\mu \nu} - A \frac{cP_1'}{8} \cos^2 \!\theta \,\varphi_{\mu} \varphi_{\nu} \nn \\
G_{\mu z} &=& g_{\mu z} - A \frac{cP_1'}{8} \cos^2 \!\theta \varphi_{\mu} \varphi_z \nn \\
G_{z z} &=& g_{z z} - A \frac{cP_1'}{8} \cos^2 \!\theta \varphi_z^2 \,\, .
\eea
Substituting $g_{\mu \nu}$, $g_{\mu z}$ and $g_{z z}$ from (\ref{8dmetric}), we find:
\bea \label{G3c}
G_{\mu \nu} &=& A \left[ \eta_{\mu \nu} + \frac{c}{4} \left( \theta_{\mu} \theta_{\nu} + \sin^2 \!\theta \,\varphi_{\mu} \varphi_{\nu} \right) \right] \nn \\
G_{\mu z} &=& A \,\frac{c}{4} \left( \theta_{\mu} \theta_z + \sin^2 \!\theta \,\varphi_{\mu} \varphi_z \right) \nn \\
G_{z z} &=& A \left[ \frac{cP_1'}{2} \left( \pd_z \rho \right)^2 + \frac{c}{4} \left( \theta_z^2 + \sin^2 \!\theta \,\varphi_z^2 \right) \right] \, .
\eea

Now we can single out the $z$ coordinate and write $\det G_{5d} = G_{zz} \det \hat{G}_{4d}$ with $\hat{G}_{4d}$ determined similarly to $G_{5d}$ in (\ref{GMN}). Therefore, we have:
\be \label{g8dexpr}
\det g_{8d} = A^3 \left( \frac{c}{4} \right)^2 \frac{cP_1'}{8} \,G_{z z} \det \hat{G}_{4d} \, ,
\ee
where $\hat{G}_{4d}$ is a 4d metric with the following components:
\be \label{G4dmunu}
\hat{G}_{\mu \nu} = G_{\mu \nu} - \frac{G_{\mu z} G_{\nu z}}{G_{z z}} \, .
\ee
In order to compute the contribution of the second term in (\ref{G4dmunu}), note that we want to expand to second order in fluctuations. Given that the classical solution corresponds to $\theta = const$, all derivatives $\theta_{\mu}$, $\theta_z$ are first order in the fluctuations. However, the classical solution for $\varphi$ depends on $z$. So, while $\varphi_{\mu}$ is first order, the derivative $\varphi_z$ has a zeroth order (i.e., classical) part. So, expanding (\ref{G4dmunu}) to quadratic order in the fluctuations, we have:
\be
\hat{G}_{\mu \nu} = G_{\mu \nu} - A \left( \frac{c}{4} \right)^2 \!\frac{\varphi_z^2}{\left[ \frac{cP_1'}{2} \left( \pd_z \rho \right)^2+ \frac{c}{4} \varphi_z^2\right]} \,\,\varphi_{\mu} \varphi_{\nu} \,\, ,
\ee
where in the coefficient of $\varphi_{\mu} \varphi_{\nu}$ we have taken the classical part of $\varphi_z$ and substituted the classical solution $\sin^2 \!\theta = 1$. Hence, to quadratic order, the determinant of $\hat{G}_{4d}$ is:
\be
- \det \hat{G}_{4d} = A^4 \left[ 1 + \frac{c}{4} \,\eta^{\mu \nu} \theta_{\mu} \theta_{\nu} + \frac{c P_1' \left( \pd_z \rho \right)^2}{2 \left[ 2 P_1' \left( \pd_z \rho \right)^2 + \varphi_z^2 \right] }  \,\eta^{\mu \nu} \varphi_{\mu} \varphi_{\nu} \right] \, .
\ee
Combining this with (\ref{g8dexpr}) and (\ref{G3c}), we finally find:
\be
\det g_{8d} = - A^8 \left( \frac{c}{4} \right)^2 \frac{cP_1'}{8} \,\left[ \frac{cP_1'}{2} \left( \pd_z \rho \right)^2 + \frac{c}{4} (\theta_z^2 + \sin^2 \!\theta \,\varphi_z^2) \right] \left( 1+ \frac{c}{4} \,\theta_{\mu}^2 + Q \,\varphi_{\mu}^2 \right) ,
\ee
where
\be \label{Q}
Q = \frac{c P_1' \left( \pd_z \rho \right)^2}{2 \left[ 2 P_1' \left( \pd_z \rho \right)^2 + \varphi_z^2 \right] } \, .
\ee
Therefore, substituting (\ref{P1p}) and the explicit form of $\pd_z \rho$ from (\ref{ChVar}), we obtain the following expression for the Lagrangian ${\cal L} = - \sqrt{-\det g_{8d}}$\,:
\be
{\cal L} = - \frac{A^4 c^2 \sqrt{\beta} \,e^{2 \rho_0}}{96} \sqrt{3(1+z^2) \!\left[4\,\theta_z^2+\sin^2 \!\theta \left(4+c\,\theta_\mu^2\right)\varphi_z^2\right]+e^{4 \rho_0}\beta\,z^2\left(4+c\,\theta_\mu^2+c\,\varphi_\mu^2\right)} \, .
\ee

\section{Estimating $\theta$ boson mass shift}
\setcounter{equation}{0}

In this Appendix we derive the result (\ref{SmallZ}) for the integral in the numerator of (\ref{Mn2}). We use the approximate wave function (\ref{interpolate}), whose failure to be an exact solution is of order $z_\Lambda^{-2}$. Hence we have:
\be
\int_{z_{\Lambda}}^{z_{\Lambda}} [\hat{\Theta}_n^0 (z)]^2 \,V_1(z) \,dz \,= \,-\frac{2}{3}\beta\,e^{4 \rho_0} I \,\, ,
\ee
where
\be
I = \int_0^{z_\Lambda} \frac{1}{\sqrt{1+z^2}}\,[J_0(M_n z)]^2 \,dz \,\, .
\ee

To evaluate $I$, let us rewrite it as:
\be \label{Atwoterms}
I = \int_0^{z_\Lambda} \frac{dz}{\sqrt{1+z^2}} \,+ \,\int_0^{z_\Lambda} \frac{1}{\sqrt{1+z^2}}\left\{[J_0(M_n z)]^2-1\right\} dz \, .
\ee
The first integral is easy to compute:
\be
I_1 \equiv \int_0^{z_{\Lambda}} \frac{1}{\sqrt{1+z^2}} \,\,dz \,= \,{\rm arcsinh} (z_{\Lambda}) \,\, .
\ee
To estimate the second integral in (\ref{Atwoterms}), let us first substitute $x=M_n z$:
\be\label{second}
I_2 \equiv \int_0^{r_n}\frac{dx}{\sqrt{M_n^2+x^2}}\left\{[J_0(x)]^2-1\right\} \, .
\ee
Although this cannot be solved analytically, we will extract now its leading contribution in a small parameter expansion. For that purpose, recall that $M_n^2 = \frac{c_0}{12} e^{4 \rho_0} \beta^{1/2} m_n^2$ with the unperturbed (i.e., without $V_1$) mass-squared being $m_n^2 \sim {\cal O} (\beta^{1/2})$ \cite{ASW}. Hence $M_n^2$ is of order $\beta$ and so can be viewed as a small parameter. Then, at zeroth order in $\beta$, we have:
\be\label{second1}
\int_0^{r_n}\frac{dx}{x}\left\{[J_0(x)]^2-1\right\}=-\frac{1}{4}r_n^2\,\,{}_3F_4 (1,1,\frac{3}{2};2,2,2,2;-r_n^2) \,\, ,
\ee
which notably is independent of $z_\Lambda$. To estimate the error in omitting $M_n$, let us rewrite $I_2$ identically as:
\be\label{second2}
I_2=\int_0^{r_n}\frac{dx}{\sqrt{M_n^2+x^2}}\left\{[J_0(x)]^2-1+\frac{x^2}{2}\right\}-\frac{1}{2}\int_0^{r_n}\frac{x^2\,dx}{\sqrt{M_n^2+x^2}} \,\, .
\ee
This is useful for the following reason. Let us consider the first integral in (\ref{second2}) and recall that for small argument $[J_0 (x)]^2 \approx 1 - \frac{1}{2} x^2 + {\cal O} (x^3)$. Then that integral does not have a logarithmic divergence in the limit $M_n^2 \rightarrow 0$ and, furthermore, its linear term in a small $M_n^2$ expansion is finite. Hence, we can obtain its contribution to ${\cal O} (M_n^2)$ by expanding the integrand in small $M_n^2$. On the other hand, the second integral in (\ref{second2}) can be performed exactly. Combining these two contributions and substituting $M_n=r_n / z_{\Lambda}$, we find:
\be
I_2=-\frac{1}{4}r_n^2\,\,{}_3F_4 (1,1,\frac{3}{2};2,2,2,2;-r_n^2) + \frac{r_n^2}{4\,z_\Lambda^2}\log(z_\Lambda)+  {\cal O} (z_\Lambda^{-2})\, .
\ee
The sum $I = I_1 + I_2$ then gives precisely (\ref{SmallZ}).

\end{document}